\documentclass[sigconf]{acmart}

\AtBeginDocument{%
  \providecommand\BibTeX{{%
    \normalfont B\kern-0.5em{\scshape i\kern-0.25em b}\kern-0.8em\TeX}}}

\setcopyright{acmcopyright}
\copyrightyear{2023}
\acmYear{2023}
\acmDOI{XXXXXXX.XXXXXXX}

\acmConference[CHI'24]{The ACM CHI conference on Human Factors in Computing Systems}{May 11--16,
  2024}{Honolulu, Hawai'i}
\acmPrice{15.00}
\acmISBN{978-1-4503-XXXX-X/18/06}

\usepackage{graphicx}
\usepackage{svg}
\usepackage{caption}
\usepackage{subcaption}
\usepackage{pdfpages}
\usepackage{hyperref}
\usepackage{lineno}
\usepackage{enumitem}

\begin{document}
\nolinenumbers

\title[Pika for Authoring Governance Policies]{Pika: Empowering Non-Programmers to Author Executable Governance Policies in Online Communities}

\author{Leijie Wang}
\email{leijiew@cs.washington.edu}
\affiliation{%
  \institution{University of Washington}
  \city{Seattle}
  \country{United States}
}

\author{Nicholas Vincent}
\email{nicholas_vincent@sfu.ca}
\affiliation{%
  \institution{Simon Fraser University}
  \city{British Columbia}
  \country{Canada}
}

\author{Julija Rukanskaitė}
\email{julija.rukanskaite@gmail.com}
\affiliation{%
  \institution{Metagov Project}
  \city{}
  \country{Sweden}
}

\author{Amy X. Zhang}
\email{axz@cs.uw.edu}
\affiliation{%
  \institution{University of Washington}
  \city{Seattle}
  \country{United States}
}

\renewcommand{\shortauthors}{Wang Leijie et al.}
\newcommand{\System}{Pika}
\newcommand\revise[1]{#1}
\newcommand\leijie[1]{#1}
\definecolor{lightblue}{HTML}{4790BA}
\newcommand{\code}[1]{\texttt{\textcolor{darkgray}{#1}}}
\newcommand{\term}[1]{\textit{#1}}
\newcommand{\dgone}{\textbf{Design Goal 1}}
\newcommand{\dgtwo}{\textbf{Design Goal 2}}


\begin{abstract}

Internet users have formed a wide array of online communities with diverse community goals and nuanced norms. However, most online platforms only offer a limited set of governance models in their software infrastructure and leave little room for customization. Consequently, technical proficiency becomes a prerequisite for online communities to build governance policies in code, excluding non-programmers from participation in designing community governance. In this paper, we present Pika, a system that empowers non-programmers to author a wide range of executable governance policies. At its core, Pika incorporates a declarative language that decomposes governance policies into modular components, thereby facilitating expressive policy authoring through a user-friendly, form-based web interface. Our user studies with 10 non-programmers and 7 programmers show that Pika can empower non-programmers to author policies approximately 2.5 times faster than programmers who author in code. We also provide insights about Pika's expressivity in supporting diverse policies online communities want.
\end{abstract}



\maketitle

\section{Introduction}





Millions of communities gather in online platforms such as Reddit, Slack, and Discord~\cite{kraut2012building}. These dynamic spaces operate based on the concept of \textit{governance}, defined as the structures, processes, and cultural norms that oversee community organization, power delegation, and the regulation of user interactions~\cite{fiesler2018reddit, preece2003history}. Governance \textit{policies} further articulate the ways in which governance is operationalized in an online community.
A carefully negotiated set of governance policies not only is crucial to the stability of online communities~\cite{ostrom2000collective}, but also impacts individual well-being and potentially broader social institutions~\cite{noveck2009wiki, matias2016going}. Reflecting their rich and nuanced community norms, online communities have thus evolved to adopt a diverse spectrum of governance policies~\cite{fiesler2018reddit, coleman2013coding, dimitri2022quadratic, muller2013work}.


However, many online platforms offer only a limited set of governance models in their software infrastructure. Predominantly, they adhere to a \textit{role-permission} governance model~\cite{zhang2020policykit, schneider2022admins} where administrators and moderators are granted more privileges than regular users, such as full authority over membership and content moderation~\cite{seering2019moderator, niederer2010wisdom}. This governance model, inherited from the earliest online platforms, is no longer a technical necessity but remains the default choice on nearly all major platforms~\cite{schneider2022admins}. While several platforms support other governance models like reputation systems~\cite{posnett2012mining, lampe2004slash} or jury systems~\cite{kou2014governance}, these often represent the sole governance option available.

\begin{figure*}[!t]
    \centering
    \includegraphics[width=\textwidth]{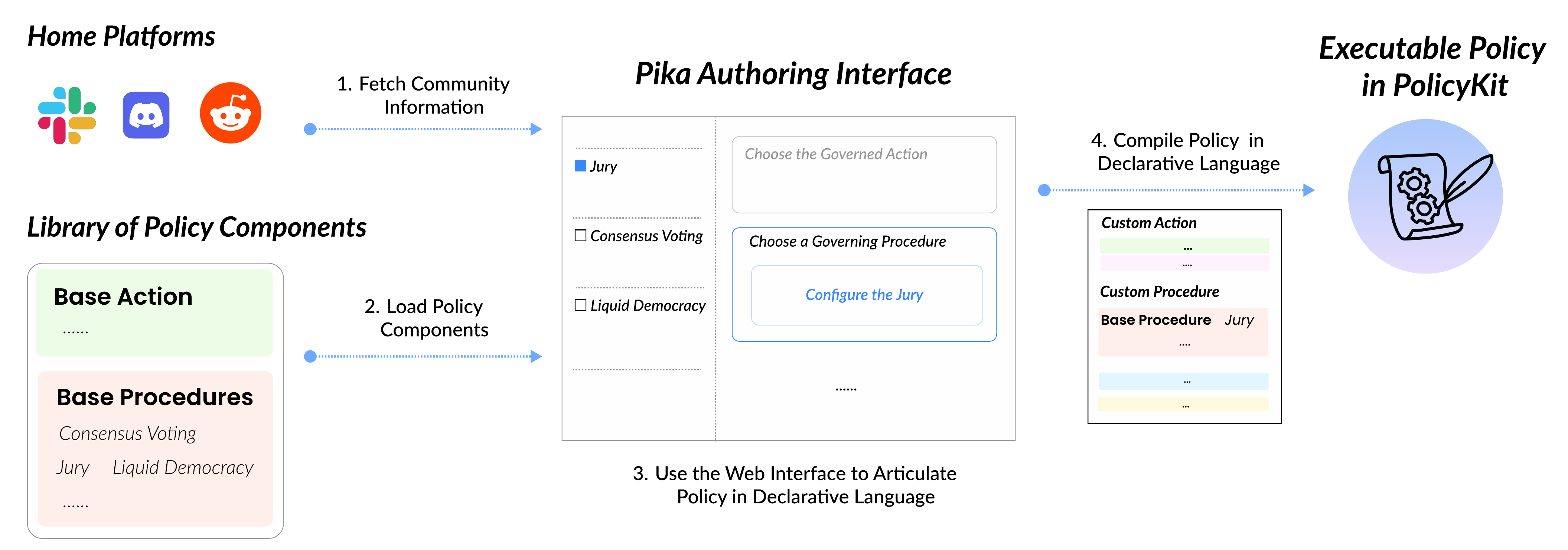}
    \caption{\textbf{\System{} Overview}. \textmd{To enable users to articulate policies in a declarative language, the Pika authoring interface fetches community information from home platforms and loads policy components from the JSON library. After the authoring process, the system generates a policy in the form of declarative language and compiles it into executable code in PolicyKit.}}
    \hfill
    \label{system implementation}
\end{figure*}

Consequently, a growing number of online communities have started \textit{implementing} governance policies that mirror their distinct norms, moving away from the default governance model. This shift is evidenced by the prevalence of bots or plugins supporting various governance policies across many online platforms, including Wikipedia~\cite{zheng2019roles}, Reddit~\cite{chandrasekharan2019crossmod}, Github~\cite{wessel2018power}, and Twitch~\cite{seering2018social}. For instance, in the popular online game Minecraft, there is an array of governance plugins, ranging from temporary bans and surveillance, to distribution of authority~\cite{frey2019emergence}. However, the variety of existing bots is still restrictive compared to the rich and ever-evolving norms and needs of online communities. This issue is further exacerbated by the limited customization options these bots offer ~\cite{chandrasekharan2019crossmod, kiene2020uses}. Communities would benefit more if they can directly author executable policies tailored to their unique community norms~\cite{de2007governance}.

Unfortunately, authoring governance policies in code requires familiarity with platform software infrastructure and programming expertise, prerequisites that are out of reach for non-programmers \cite{long2017could}. More recently, PolicyKit has emerged as a software infrastructure that enables online communities to implement executable governance policies without worrying about integrating them with platform software~\cite{zhang2020policykit}. Nevertheless, proficiency in Python programming and familiarity with the PolicyKit documentation remain fundamental requirements for authoring policies in PolicyKit.

If authoring code is the primary avenue for building governance policies, then technical proficiency becomes an insurmountable barrier to designing community governance, a value-laden process where there should be no barrier for anyone invested in community governance~\cite{ostrom2000collective, viegas2007hidden}. This has several implications. First, the majority of communities without technical developers have to live with the default governance model and available bots even if they desire other forms of governance. Second, it can also lead to the reinforcement of technocratic autocracies within an online community. Because they cannot comprehend and implement governance policies in code, non-technical members are excluded from auditing and authoring governance policies~\cite{hirschman1970exit}.

In this work, we present \System{}, a novel system that empowers non-programmers to create a broad range of governance policies that are directly executable on their home platforms (Figure \ref{system implementation}). Our system consists of three parts.

\begin{itemize}[leftmargin=0.3cm]

    \item \textbf{A declarative language}: At its core is a declarative language that decomposes governance policies into modular policy components (Figure~\ref{overview of data models}), thereby enabling expressive policy authoring for non-programmers. For example, while online communities use various governance procedures to make decisions, a core set of procedures are commonly used (e.g., jury, consensus voting). Hence, we can separate the customization of these procedures (e.g., notifying individuals who haven't voted) from the procedures themselves.

    \item \textbf{A library of policy components}: The modular design makes it possible to have a library of policy components and flexibly add new ones. We implemented a library of policy components (Figure \ref{implemented procedure}) to support the authoring of common policies used by online communities.

    \item \textbf{An authoring interface}: Finally, we built a user-friendly, form-based web interface (Figure \ref{web interface}) for non-programmers to author policies, which are then compiled into executable policies.
    
\end{itemize}

We conducted a user study with 10 non-programmers who had experience in community governance to assess \System{}'s usability and expressivity. They were asked to author policies using \System{} in two distinct governance scenarios. 
Additionally, we recruited 7 programmers who were interested in community governance to compare the efficiency of \System{} with a baseline of programming policies using PolicyKit. These programmers were required to undertake the same tasks as the non-programmers, with an added task of programming one randomly chosen policy. 
Our findings revealed that non-programmers were capable of authoring governance policies via \System{} approximately 2.5 times faster than programmers coding the same set of policies.
We also found that \System{} was rated by programmers as significantly more usable than the baseline, and that non-programmers also found \System{} usable. Finally, we also gained insights about \System{}'s expressivity, finding that non-programmer participants can use our declarative language to articulate the majority of policies they proposed. 



\section{Related Work}
\leijie{In this section, we begin by examining the tension between the diverse governance needs of online communities and the limited governance models offered by online platforms. In response to this, online communities have started to build technical governance tooling. This trend inadvertently excludes non-programmers from participating in designing and auditing community governance. To address this, we draw inspirations from the end-user programming literature to empower non-programmers to articulate policies using a declarative language.}

\subsection{Governance in of Online Communities}
Online community governance refers to the systems, structures, processes, and cultural norms that guide, manage, and oversee decision-making within communities~\cite{preece2003history, williamson1999strategy}.
Within this broad scope, governance \textit{policies} describe how governance becomes operationalized in online communities. Policies in this context are neither traditional public polices or laws, nor Terms of Service externally imposed by platforms. Instead they articulate both informal norms and formal rules by which communities organize members, delegate power, and regulate user interactions~\cite{fiesler2018reddit, butler2008don}. 
Policies play a critical role in combating online harassment~\cite{chandrasekharan2019crossmod}, resolving disputes~\cite{im2018deliberation, kittur2007he}, and even fostering offline political coordination~\cite{aragon2017deliberative, kling2015voting}. A carefully negotiated set of policies is vital to the long-term stability and survival of online communities~\cite{ostrom2000collective}.

Online communities have embraced a diverse spectrum of governance policies. The most studied example is Wikipedia~\cite{muller2013work, forte2008scaling, lovink2012critical} where more than 800 pages are dedicated to articulating various policies and guidelines~\cite{zheng2019roles}. Policies vary greatly to accommodate diverse norms and needs of individual communities, even when located on the same platform~\cite{fiesler2018reddit}. 
Moreover, many online communities learn from offline governance practices. For instance, Wikipedia has an elected board that functions as a judiciary body, tasked with policy interpretation and formal arbitration~\cite{forte2008scaling}. Similarly, the Debian Project is governed by a liberal-democratic constitution that meticulously prescribes the procedures for administrator election and community membership~\cite{coleman2013coding}. More recently, tech-savvy communities have begun to explore innovative governance practices, such as quadratic voting~\cite{dimitri2022quadratic} and liquid democracy~\cite{hardt2015google}.

Despite the evident need for a variety of governance policies in online communities, online platforms offer a limited set of governance models in their software infrastructure. Predominantly, these platforms follow a \textit{role-permission} governance model~\cite{zhang2020policykit, schneider2022admins}. In this model, administrators and moderators are granted more privileges than regular users, such as full authority over membership and content moderation~\cite{seering2019moderator, niederer2010wisdom}. It originated from the earliest online communities, such as BBSes~\cite{malloy2016origins}, Usenet~\cite{spencer1998managing}, and email-list platforms, where one technical administrator is required to operate the community server. As online communities transitioned from self-hosted servers to a small number of commercial platforms, this governance model is no longer a technical necessity but remains the default choice~\cite{schneider2022admins}. While several platforms support other governance models like reputation systems~\cite{posnett2012mining, lampe2004slash} or jury systems~\cite{kou2014governance}, these often represent the sole governance option available.

\subsection{Tools to Support Governance Policy Authoring and Execution}
Nevertheless, online communities have started to move away from the default governance model via several approaches, a journey that can entail significant efforts and resources. Many online communities \textit{manually} maintain and execute governance policies that are either implicitly recognized or expressly documented~\cite{fiesler2018reddit, forte2009decentralization}. However, the manual execution of policies demands substantial human efforts and becomes increasingly infeasible as the community grows~\cite{im2018deliberation, mnookin2017virtual}. This has led to a noticeable shift towards technical governance tooling~\cite{muller2013work}, characterized by the increasing prevalence of bots or plugins that encapsulate policies into code and automatically execute them~\cite{halfaker2012bots}.
Deployed across various online platforms, including Reddit~\cite{chandrasekharan2019crossmod}, GitHub~\cite{wessel2018power}, and Twitch~\cite{seering2018social}, these bots are used to perform repetitive and laborious tasks regarding content moderation and newcomer onboarding. 
In Wikipedia, a significant fraction of the 1,601 registered bots undertake the duty of upholding community standards and regulating user behaviors~\cite{zheng2019roles, geiger2010work}. In Minecraft, a popular online game, communities have access to an array of governance plugins, ranging from temporary bans and surveillance to the distribution of authority~\cite{frey2019emergence}. 

However, compared with the rich and ever-evolving norms of online communities, the variety of existing bots is still restrictive. It is evidenced by the complaints that many basic features of offline governance legacies are not available on online platforms~\cite{schneider2021modular}.
The issue is exacerbated by the limited customization options these bots offer. For instance, Automod on Reddit allows moderators to author only simple rules using regular expressions for removing unwanted messages, offering limited flexibility for alternative moderation actions beyond message removal~\cite{chandrasekharan2019crossmod}. As communities' perception of their desired governance model continually evolves over time~\cite{de2007governance}, there is a growing demand for governance policies specifically tailored to their community norms.

Authoring governance policies in code requires familiarity with platform software infrastructure and programming expertise, prerequisites that are out of reach for non-programmers. Long et al. documented the challenges non-programmers have when requesting or creating Reddit bots~\cite{long2017could}. They grapple not only with the complexities of the Reddit API and programming but also with assessing the feasibility of their bot requests. Recently, PolicyKit has enabled online communities to articulate executable governance policies in code without worrying about integrating them with platform APIs~\cite{zhang2020policykit}. Nevertheless, proficiency in programming remains a fundamental requirement for authoring policies in PolicyKit.

If authoring code becomes the primary avenue for building governance policies, then technical proficiency is an insurmountable barrier to participation in designing community governance. This trend has two implications. First, communities without technical developers must adapt to the default governance model a platform offers even when they prefer other forms of governance. They are also obliged to find ways to work with bots that may have been designed for communities with slightly different needs. Second, non-programmer members are inadvertently excluded from discussion, deliberation, and articulation of governance policies in code, a value-laden process where there should be no barrier to participate~\cite{ostrom2000collective, viegas2007hidden, geiger2014bots}. 
Consequently, non-technical community leaders are forced to depend on skilled developers for policy implementation, creating a potential disconnect~\cite{o2007emergence}. Meanwhile, non-programmer members struggle to comprehend governance policies in code, which, in turn, hampers their ability to audit these policies and effectively voice their concerns~\cite{hirschman1970exit}.

\subsection{End-User Programming}
End-user programming enables non-programmers to tailor computer technologies to meet their personal or professional needs~\cite{scaffidi2005estimating}, such as analyzing data with spreadsheets~\cite{nardi1990spreadsheet, hermans2011supporting}, authoring web pages~\cite{verou2016mavo, alrashed2022wikxhibit}, or connecting Internet of Things (IoT) devices~\cite{dey2006icap, akiki2017visual}.
On online platforms, end-user programming systems such as IFTTT (short for ``\textit{if-this-then-that}'')~\cite{ur2016trigger, mi2017empirical} and Zapier enable everyday users to automate tasks, including facilitating email conversations~\cite{kokkalis2017myriadhub}, and collecting community responses. 
Some online platforms also build tools that enable community moderators to define rules and filters to remove unwanted content~\cite{jhaver2019human, kiene2020uses}. For instance, moderators can configure simple rules using regular expressions to filter unwanted messages~\cite{jhaver2019human, jhaver2022designing}, or create simple conditional statements to trigger different moderation actions~\cite{chandrasekharan2019crossmod}. 
However, these end-user programming systems offer limited expressivity and are restricted to a narrow scope of community governance policies. In comparison, our system empowers non-programmers to author an extensive array of governance policies.

End-user programming has been enabled through a variety of methods, including programming by demonstration~\cite{dey2004cappella, li2017sugilite}, 
wizard-based interaction~\cite{castelli2017happened, schobel2016end}, and natural language programming~\cite{van2010atomate, gulwani2014nlyze}. An overview of these approaches is provided by Myers et al. \cite{myers2006invited} and Barricelli et al.~\cite{barricelli2019end} Central to many approaches is the use of declarative languages. In contrast to imperative languages that specify ``how'' to achieve a particular outcome, declarative languages focus on the ``what''---the desired outcome---without detailing the exact steps to reach it~\cite{satyanarayan2016vega}. Given their nature, declarative languages can often be paired with visual interfaces, allowing users to interact with higher-level constructs than code. 

One notable example of declarative languages is \textit{trigger-action programming}, in which a user associates a trigger with an action, such that the action is automatically executed when the trigger event occurs. This programming model empowers non-programmers to express automated tasks on smart devices and online platforms ~\cite{dey2006icap, brush2011home}. But researchers have also criticized its over-simplification and limited expressivity~\cite{huang2015supporting}. For instance, IFTTT does not allow users to create rules with conjunctions of multiple triggers~\cite{ur2014practical}. We draw design inspiration from trigger-action programming, as a subset of governance policies can be considered specialized automated tasks. Nevertheless, rather than simply conditioning actions on triggers, governance policies often require input from community members and need a more expressive declarative language.

\subsection{Abstractions for Describing Governance in PolicyKit}
\label{policykit}
We aim to enable end-users to author a wide range of governance policies that are executable on their home platform of choice. 
Since PolicyKit obviates the need to work directly with platform APIs~\cite{zhang2020policykit}, our system builds on top of the open-source PolicyKit infrastructure. 
As a result, the abstractions underlying our policy authoring process further extend PolicyKit's abstractions of \textbf{actions} and \textbf{procedures}. 

These two abstractions stem from the observation that, across myriad governance policies, we can describe the speciﬁc behavior that is being proposed separately from the rules being used to determine whether that behavior is allowed. Specifically, an action describes an event that can occur within a community and is typically first proposed by a community member. In contrast, a procedure is a structured process that determines whether a proposed action should pass or fail. 
For example, a policy might state that ``\textit{renaming of any channel must first be approved by a random jury of members}''. An example of an action governed by this policy might be, ``\textit{Bob tries to rename the channel \#general}'', while the procedure would detail the process of selecting, convening, and determining the outcome of a jury.
We break down PolicyKit's abstractions of actions and procedures into modular components, which form the basis of our declarative language and end-user authoring tool.

In addition to leveraging PolicyKit's abstractions, we use its policy engine infrastructure, which perpetually monitors actions, checks them against written policies, orchestrates any asynchronous voting that takes place as required by policies, and and executes approved actions on the platform.
Policies written in our declarative language are compiled into Python code readable by PolicyKit.
This enables our system to focus on lowering barriers to policy authoring, as PolicyKit handles execution.

\section{Overview of \System{}}
We introduce \System{}, a novel system that empowers non-programmers to efficiently author a broad range of executable governance policies using a form-based web interface. Our system is motivated by the following two design goals. 

\textbf{Design Goal I: Empower users to author policies \textit{without programming.}} Since technical proficiency is a prerequisite for building governance policies, non-technical communities are excluded from articulating their policies that adapt to their unique culture and norms. 
Even if a community has the technical expertise to implement policies, it is also important to ensure that non-programmer community members can contribute to and evaluate governance policies.

\textbf{Design Goal II: Enable \textit{expressive} authoring of policies.} Nowadays, online communities enforce a variety of governance policies according to their diverse needs and norms. 
This motivates us to ensure that our system provides non-programmers with nearly the same level of expressivity and flexibility as traditional coding.

However, a fundamental tension exists between two design goals. \revise{An end-user-friendly language often employs a higher level of abstraction, constraining non-programmers to author policies with nuanced expressiveness~\cite{barricelli2019end}.}
For example, one potential approach is to enable the sharing of policy templates in code~\cite{castelli2017happened}. 
While non-programmers can easily configure a policy template, they may even struggle to replace the governed action with another one. 


To resolve this tension, we draw inspiration from prior systems such as IFTTT~\cite{ur2016trigger} and Vega-lite~\cite{satyanarayan2016vega} to build a declarative language with a modular and decomposable grammar for governance policies. Compared to policy templates, our declarative language provides non-programmers with greater expressivity to articulate their desired policies. Further, policies in a declarative language can be easily embedded in a form-based interface. It can also lower the threshold for non-programmers to understand policies and thus increase the transparency of community governance. Specifically, \System{} consists of the following three components. 

 \begin{itemize}[leftmargin=0.3cm]
    \item A declarative language that decomposes a policy into modular policy components with fully specified attributes,
    \item A library of policy components to support the authoring of commonly used policies, and
    \item A form-based authoring interface that guides non-programmers to articulate a policy using declarative language, and compiles the authored policy into an executable policy in PolicyKit.
\end{itemize}


\begin{figure*}[!h]
    \centering
    \includegraphics[width=\textwidth]{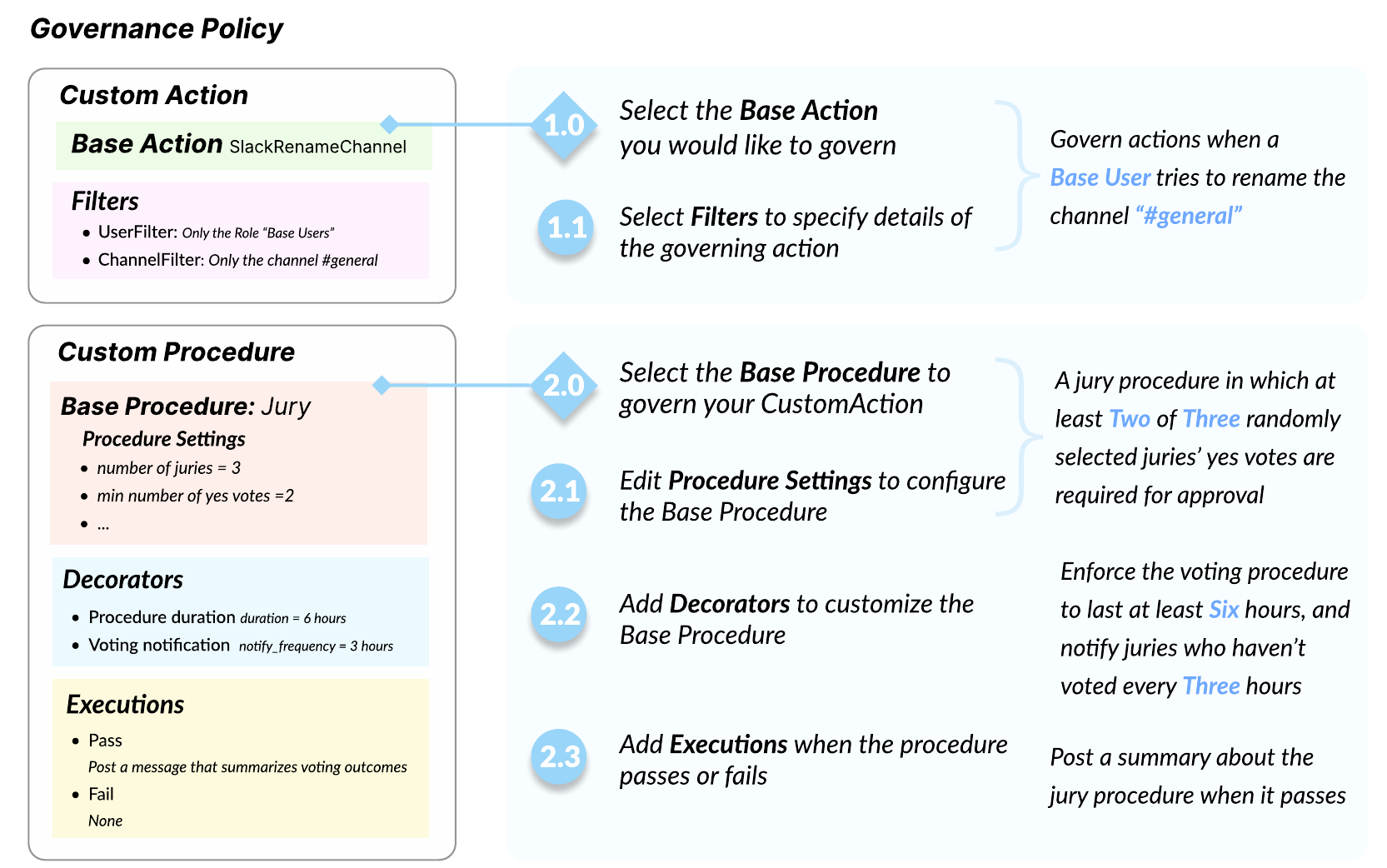}
    \caption{\textbf{An Overview of Data Models}. \textmd{Extending PolicyKit's abstractions of actions and procedures, we further decompose a \term{custom action} into a \term{base action} and \term{filters}, as well as a \term{custom procedure} into a \term{base procedure}, \term{decorators}, and \term{executions}. Such decomposition also grants non-programmers more expressivity in articulating a policy. Here we present approaches for users to articulate their governed action and governing procedure in our declarative language.}}
    \hfill
    \label{overview of data models}
\end{figure*}

\begin{figure}
    \centering
    \includegraphics[width=\columnwidth]{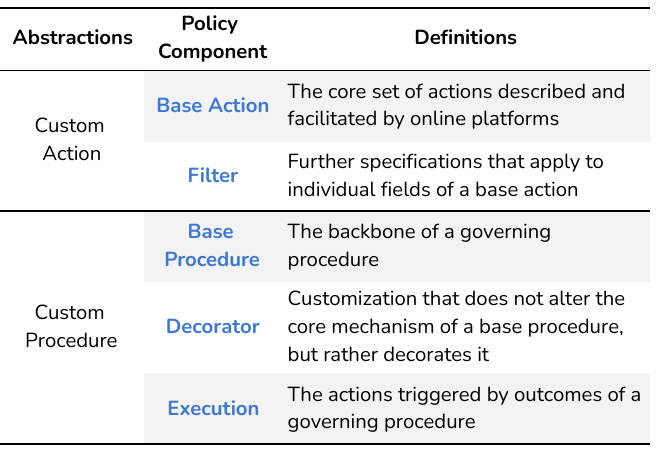}
    \caption{\leijie{\textbf{Definitions of Policy Components} }}
    \hfill
    \label{policy_component_def}
\end{figure}

\section{Design of \System{}'s declarative language}
While PolicyKit offers valuable abstractions of actions and procedures for policy formulation~\cite{zhang2020policykit}, our design goals compel us to refine these high-level abstractions into a modular grammar. In this section, we are driven to answer the question: what are the components that make up a policy, and how can they be pieced together?
We first introduce the central concepts of \term{custom actions} and \term{custom procedures}, and then delve deeper into the grammar of our declarative language. Following this, we describe how to compile this language into executable code for use with governance tooling like PolicyKit. Finally, we explain our efforts to build a shared library of policy components. As this section describes specific design choices using unique terms and platform-specific names, we use several fonts to make this section easier to follow: we emphasize different \term{terms} (e.g., \term{custom actions}) and direct references to \code{PlatformSpecificNames} (e.g., \code{SlackRenameChannel}).


\subsection{Actions and Procedures}

As previously discussed in section \ref{policykit}, the two main abstractions within PolicyKit are \textbf{actions} and \textbf{procedures}. An action refers to an event initiated by a community member within a community, while a procedure is a structured process that evaluates the potential approval or rejection of a proposed action. Here, we extend these two concepts to \term{custom actions} and \term{custom procedures}, highlighting the customization options for end-users. In the following, we will explain how we break them down into a series of policy components. The concept \textbf{policy component} is our broad term for capturing all the different modular elements that constitute a policy, namely: \term{base actions}, \term{filters}, \term{base procedures}, \term{decorators}, and \term{executions}.

\subsubsection{\textbf{Custom Action}} Consider a scenario where a policy author wishes to govern the action when a \code{Base User} renames the \code{\#general} channel in a Slack team. As illustrated in Figure \ref{overview of data models}, this can be articulated as a \term{custom action} in our declarative language, comprising the following two policy components.

\begin{itemize}[leftmargin=0.3cm]
\item
\textbf{Base Action}: First, the policy author should determine the \term{base action} they want to govern. \term{Base actions} are defined as\textit{the core set of actions described and facilitated by online platforms}. The unique technical affordances of each online platform delineate a limited set of \term{base actions}, such as renaming a channel on Slack or creating a post on Reddit. In practice, the set of \term{base actions} can be determined by a platform's API and event listening offerings. Hence, regardless of the action a policy author wants to govern, they can always select one of these \term{base actions} as a starting point. In this scenario, the choice would be \code{SlackRenameChannel}.

\item
\textbf{Filters}: In the next step, the policy author should select and configure \term{filters} to provide more detailed descriptions of their governing action. Here, \term{filters} are defined as \textit{further specifications that apply to individual fields of a \term{base action}}. For the \term{base action} \code{SlackRenameChannel}, authors can specify its action fields: \code{initiator}, \code{channel}, and \code{new\_name}. As the governing action in this case restricts the initiator to have the role of \code{Base User} and the channel subject to renaming to \code{\#general}, the author should select the corresponding \term{filters} for these two action fields respectively. As an intentional design choice, we restrict each \term{filter} to specify one action field and assume that multiple \term{filters} for the same \term{base action} are combined using the logical operator "\code{and}"\footnote{If authors want to govern actions that satisfy this or that condition, they can set up two respective policies.}. This approach enables us to focus on designing \term{filters} independently without worrying about how multiple \term{filters} might interact with each other. 
\end{itemize}

Figure \ref{declarative langauge}.(a) presents the resultant declarative language for this \term{custom action}.

 \subsubsection{\textbf{Custom Procedure}} Figure \ref{overview of data models} then illustrates how the policy author proceeds to articulate the \term{custom procedure} in our declarative language. They first select \code{Jury} as the \term{base procedure}, which is defined as \textit{the backbone of a governing procedure}. This concept stems from the observation that, despite the myriad of governing procedures found on online platforms, a core set of governing procedures are commonly used. They could be traditional ones such as consensus voting, a jury, or dictatorship, or more sophisticated ones in tech-savvy communities, such as quadratic voting~\cite{dimitri2022quadratic} or liquid democracy~\cite{hardt2015google}. While the specifics of these procedures may vary, they can be reduced to a limited set of \term{base procedures} that encapsulate their core mechanism. On the other hand, even a simple consensus voting procedure can have a variety of nuances in practice. To facilitate greater expressivity, our declarative language enables authors to customize a \term{base procedure} in three distinct ways.

\begin{itemize}[leftmargin=0.3cm]
    \item \textbf{Settings of Base Procedures}: First, after specifying a \term{base procedure}, policy authors can configure its \term{settings} that represent its important parameters. For instance, the number of jurors and the threshold of affirmative votes dictate how a jury procedure operates; authors from a small community might set the required number of jurors to three. Each \term{base procedure} define a list of \term{settings} as configuration interfaces, enabling initial customization.     
    \item \textbf{Decorators}: Second, policy authors might want to add additional customization to their selected \term{base procedure}, such as enforcing time restrictions on voting or notifying people who haven't voted. Such customization does not alter the core mechanism of a \term{base procedure}, but rather \term{decorates} it. Analogous to decorators in Python, \term{decorators} adjust the behavior of a \term{base procedure} without tampering with how it determines the final result.
    
    \item \textbf{Executions}: Finally, policy authors might want to have additional actions occur following the outcome of a procedure. For instance, in addition to executing the governed action when the procedure passes, they might notify community members about the voting outcomes. We term \textit{the actions triggered by outcomes of a governing procedure} as \term{Executions}. Similar to \term{base actions}, the set of \term{executions} available is determined by platform affordances. However, they are conceptually different: \term{executions} describes actions that are to happen, while \term{base actions} refer to actions that have already happened. 
\end{itemize}

\begin{figure*}
    \centering
    \includegraphics[width=\textwidth]{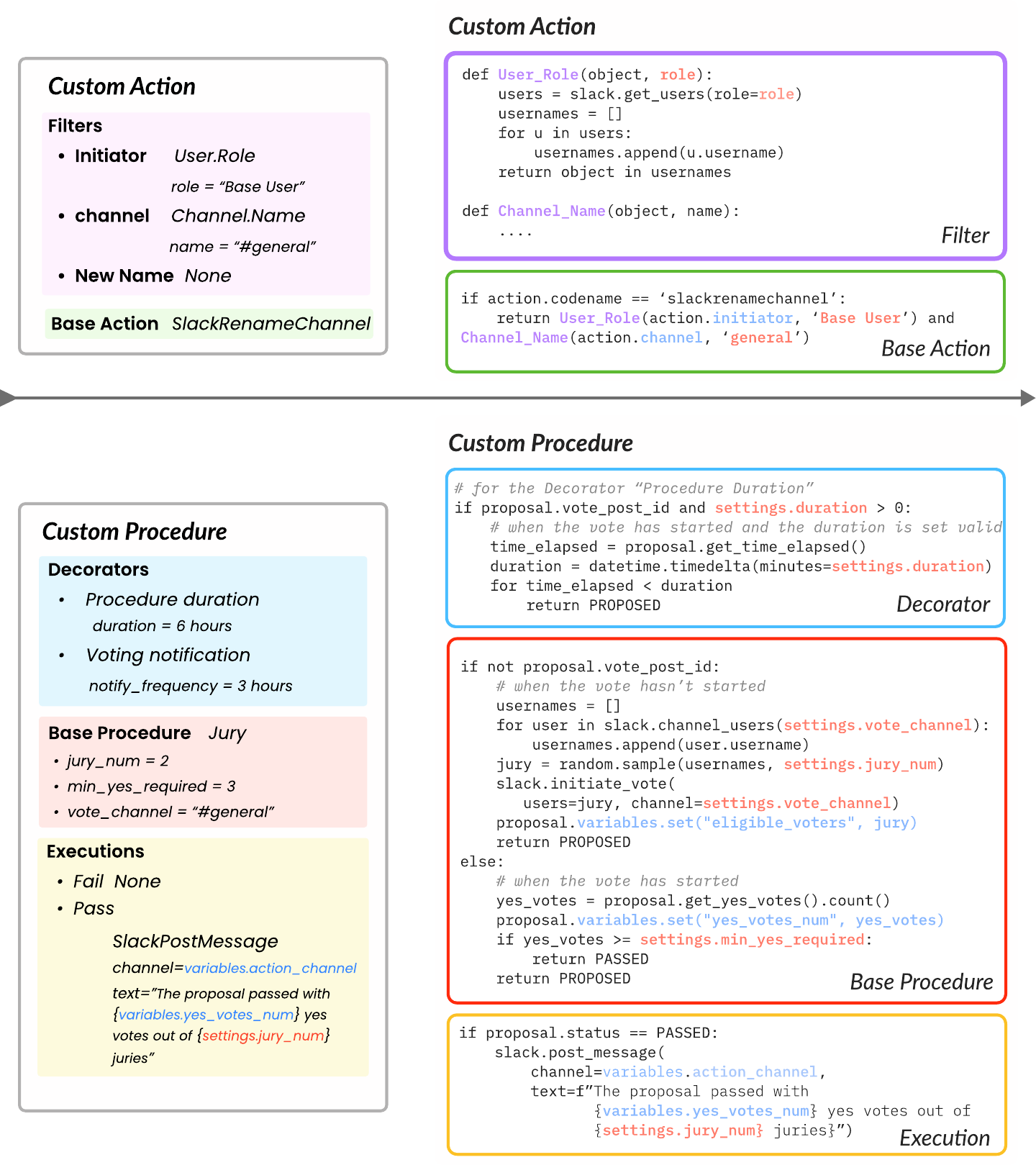}
    \caption{\textbf{(left) A Policy Written in the Declarative Language}. \textmd{This policy is corresponding to users' configuration in Figure \ref{overview of data models}. For clarity, a boxed view is used instead of the actual JSON format. Note the references to \term{variables} (\textcolor{blue}{blue}) and \term{settings} (\textcolor{red}{red}) in the \code{channel} and \code{text} fields of the \term{execution} \code{SlackPostMessage}. This \term{execution} is set to happen in the channel where the \term{custom action} happens and its message refers to the number of jurors and the number of final yes votes from the \term{base procedure}}.\\
    \textbf{(right) The Corresponding but Simplified Executable Python Code in PolicyKit}. \textmd{We generate the code by linking and formatting the code snippet for each mentioned policy component according to PolicyKit's requirements. The declarative language components and their corresponding code segments are color-coded for easy correlation. Similar to decorators in Python, the code of \term{decorators} is prepended to the code of the \term{base procedure}.}}
    \hfill
    \label{declarative langauge}
    \label{executable code}
\end{figure*}

Figure \ref{declarative langauge}.(c) presents the resultant declarative language for this \term{custom procedure}.

\begin{figure*}
    \centering
    \includegraphics[width=\textwidth]{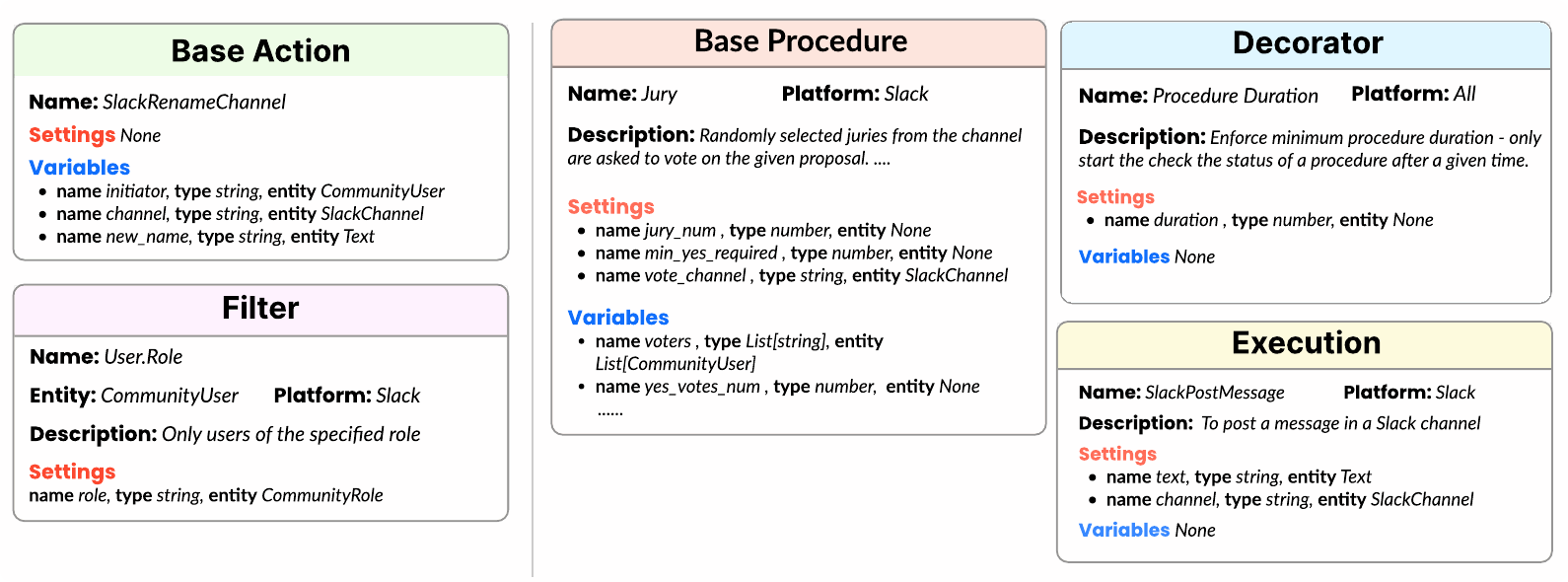}
    \caption{\textbf{Details of policy components}. \textmd{Policy components are designed in a similar pattern: the \code{name} serves as its unique identifier, while the \code{description} offers a detailed explanation for end-users. The \term{settings} stipulate the necessary user input, complemented by \term{variables} that facilitate cross-component references. Each policy component is also paired with a respective code snippet used to compile policies in the declarative language into executable code in PolicyKit.}}
    \hfill
    \label{component library}
\end{figure*}

\subsection{Compiling Declarative Language into Executable Code}

As illustrated in Figure \ref{declarative langauge}, we built a compiler to translate a policy in the declarative language (left) to the executable Python code in PolicyKit (right). PolicyKit provides access to platform functions and requires a series of Python functions with expected inputs and outputs. Hence, we compile the code snippets associated with each policy component into the respective functions. Similar to decorators in Python, code of selected \term{decorators} is prepended to the code of the selected \term{base procedure}. This is to ensure \term{decorators} override the behavior of the \term{base procedure}.

\subsection{\textbf{Articulating
Policy Components}}

In \System{}, policy components act as a bridge connecting programmers and non-programmers: programmers are expected to implement a policy component in code and outline its customization interfaces, which, in turn, allows non-programmers to effortlessly engage in the policy authoring process.
As online communities often use a similar set of policy components, this approach can substantially reduce the workload involved in rebuilding them from scratch and promote the sharing of policy components with non-technical communities. Therefore, it is vital to maintain clearly defined structures for each category of policy components.

As illustrated in Figure \ref{component library}, all policy components follow a consistent structure, represented using JSON fields and accompanied by a corresponding code snippet. 
Specifically, every policy component is characterized by the following JSON fields: an identifier (typically \code{name}) and \code{description} that describe each policy component to end-users. Moreover, as elaborated below, it is imperative for each policy component to define a list of \term{settings} and \term{variables}.

\subsubsection{\textbf{Settings of Policy Components}} \term{Settings} outline potential ways that users can configure a policy component. Beyond \term{settings} of a \term{base procedure} that we have previously addressed, other policy components also have their respective \term{settings}. For example, as depicted in Figure \ref{component library}, a user can specify the required \code{role} for the \term{filter} \code{User.Role}. They can also set the \code{duration} for a \term{decorator} that enforces procedure duration or personalize the \code{text} of the \term{execution} \code{SlackPostMessage}. \term{Settings} provide a useful abstraction of a policy component so that end-users no longer need to understand the intricate details of its implementation (\dgone{}).

While the name of a \term{setting} may imply expected values from a policy author, it is critical to provide more detailed descriptions for each \term{setting} to facilitate user choices and validate their inputs. While categorizing a \term{setting} based on its data type (e.g., \code{number}, \code{string}) might seem natural from a technical standpoint, this approach falls short in achieving both aforementioned objectives. For instance, the \term{setting} \code{voting\_channel} is represented in the backend as a string of random alphanumeric (i.e., channel IDs) rather than its more intuitive, readable names (e.g., \code{\#general}). Identifying and then inputting the correct value for this \term{setting} could be a daunting task for non-programmers. Moreover, we cannot verify the validity of users input simply through data type information. 

Therefore, we further enriched the description of a \term{setting} by incorporating details about its linked entity, if applicable. Some \term{settings} reflect entities within a community, such as \code{CommunityUser} and \code{SlackChannel}. While data types dictate the representation and storage of a \term{setting} in code, entities are more recognizable and intuitive to end-users, facilitating a smoother configuration process (\dgone{}). For instance, as the \term{setting} \code{voting channel} permits values of the entity \code{SlackChannel}, we could present users with a predefined list of Slack channels fetched from the community.

\subsubsection{\textbf{Variables of Policy Components}} \term{Variables} of a policy component provide useful information for references across components. While a modular grammar of policies is more accessible and expressive for end-users' articulation, it also presents a challenge to connect independently implemented policy components. This could inadvertently limit end-users' ability to fully express certain policies (\dgtwo{}). For instance, users may want to cast a vote in the channel subject to renaming or send thank-you messages to the selected jurors after the procedure ends. Such policies are not possible if users cannot reference information about the channel subject to renaming or selected jurors across policy components. To overcome this challenge, we stipulate that each policy component should define a list of \term{variables} for references across components. Unlike \term{settings}, \term{variables} are not open to user configuration but instead, are results of user configuration. Examples of \term{variables} include the \code{channel} of the \term{base action} \code{SlackRenameChannel}, or the selected juries of the \term{base procedure} \code{Jury}. 

As users continuously add new policy components in the authoring process, we expect a considerable number of \term{variables} in the authoring environment. This can potentially overwhelm users when they attempt to reference a \term{variable} for a specific \term{setting}, struggling to determine which \term{variable} constitutes a valid input for this \term{setting}. To reducing the cognitive burden on end-users in such scenarios, entities of \term{variables} helps narrow down potential \term{variables} by filtering \term{variables} of the same entity as the \term{setting} (\dgone{}).

\subsection{A Library of Policy Components}
\label{libraryofcomponents}
While our declarative language offers expressive grammar to describe a policy, it will only become useful to non-programmers when coupled with a diverse library of policy components as building blocks. As a starting point, we implemented a library of policy components to support a decent number of common policies used by online communities. The modular design of our declarative language also leaves room for programmers to add new policy components--it simply involves adding them to the JSON library. We now introduce the library of policy components in more detail.

\begin{figure}[!b]
    \centering
    \includegraphics[width=\columnwidth]{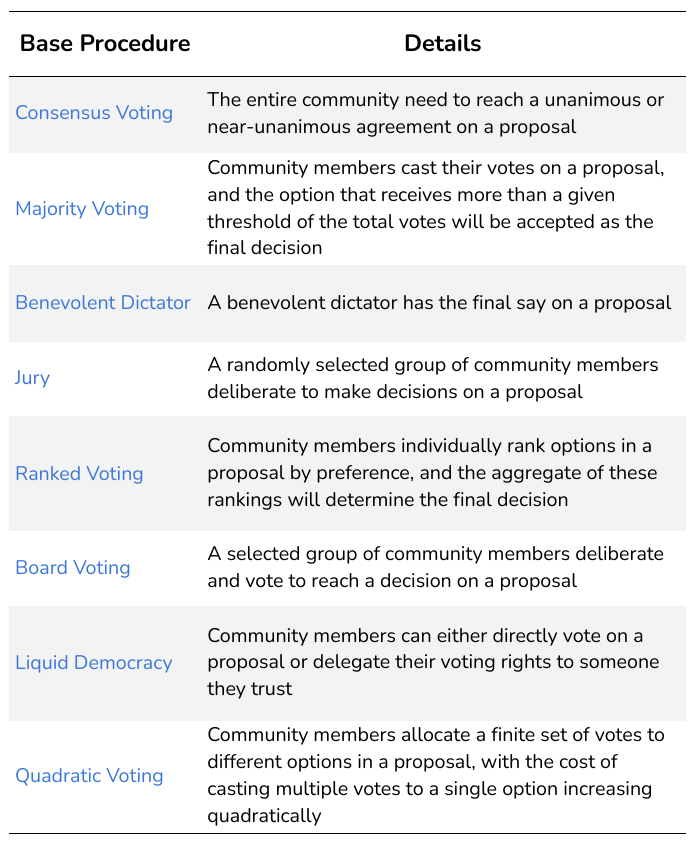}
    \caption{\textbf{A list of implemented \term{base procedures}}.}
    \hfill
    \label{implemented procedure}
\end{figure}

\begin{itemize}
    \item \textbf{Base action} and \textbf{Executions}. We support a set of common \term{base actions} and \term{executions} on Slack (e.g., renaming a Slack channel, inviting a user to a Slack channel) and those more closely related to community governance (e.g., granting a role to a user, editing community documents).

    \item \textbf{Filters}. We implemented nearly 20 \term{filters} to support further specification of a variety of entities (e.g., \code{CommunityUser}, \code{Text}, \code{Timestamp}, \code{SlackChannel}). 

    \item \textbf{Base procedure}. As listed in Figure ~\ref{implemented procedure}, we implemented a series of \term{base procedures} that represent the commonly used decision-making procedures by online communities, ranging from \code{Consensus Voting}, \code{Benevolent Dictator} to more complex procedures such as \code{Ranked Voting}, \code{Quadratic Voting}, and \code{Liquid Democracy}.

    \item  \textbf{Decorators}. We focused on the customization of voting procedures and implemented \term{decorators} that notify people who have not voted, require all eligible voters to vote, and delay voting checks. 
    
\end{itemize}

It's important to highlight that our objective isn't to narrow down each kind of policy component to the most succinct set possible. Rather, our focus is on fostering a straightforward comprehension for non-programmers. 
Hence, we allowed some implemented policy components to share a similar core mechanism. For example, configuring the threshold of a majority vote to be nearly 100\% effectively transforms it into a consensus voting procedure. In this way, non-programmers can select procedures they are familiar with, without the necessity of delving into the complex conceptual variances in each category of policy components.

\section{Web Interface}

\subsection{Web Interface}
\begin{figure*}
    \centering
    \includegraphics[width=\textwidth]{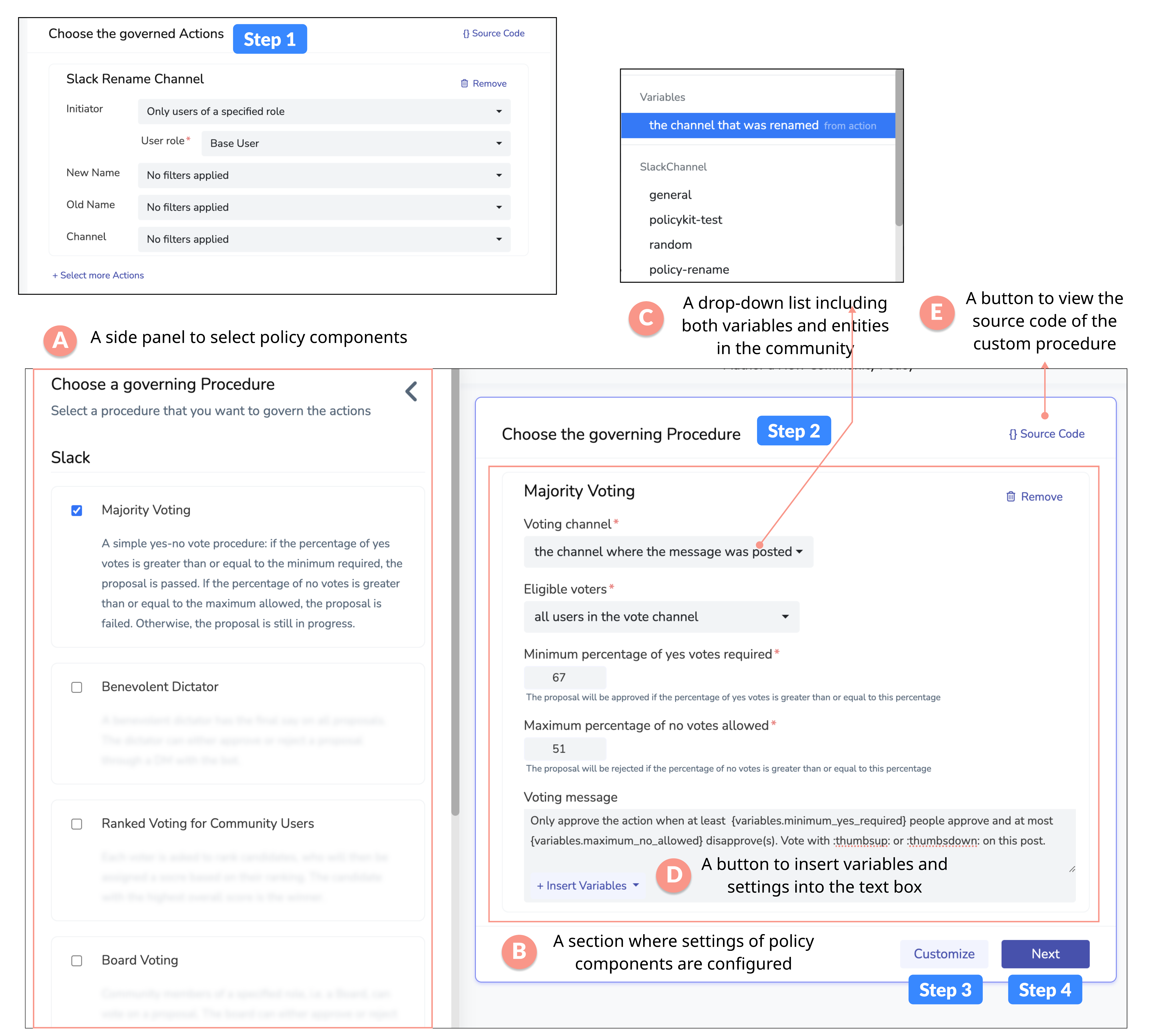}
    \caption{\textbf{The Web Interface of \System{}}. \textmd{This interface guides users through the policy articulation process in the following steps, as indicated by the numbered steps in the figure: (1) configure the \term{custom action} by selecting \term{base actions} and \term{filters}, (2) choose and configure the \term{base procedure}, (3) choose the \term{decorators}, and (4) choose the \term{executions}. While technical terms are used to explain these steps here, the web interface employs user-friendly language for non-programmers. \hfill\\
    Additionally, the figure highlights several noteworthy interactions, marked by letters. (A) The \leijie{side panel} allows users to select and scrutinize various options for each policy component (e.g., \term{base actions}, \term{base procedures}, \term{decorators}). (B) This section aids users in configuring the \term{settings} of each policy component, presenting input boxes with validations that correspond to each \term{setting}'s data type and entity. (C) For \term{settings} corresponding to a specific entity, such as \code{voting channel}, a drop-down menu that includes both matching \term{variables} and entities in the community facilitates easier input than a standard input box (D) The "Insert Variables" button enables users to reference \term{variables} and \term{settings} directly in the text. (E) Users with a basic programming background can access and view the source code for each policy component.}}
    \hfill
    \label{web interface}
\end{figure*}

Incorporating both a declarative language and access to the library of policy components, our web interface serves as an important tool for empowering non-programmers to successfully author executable policies. Following the steps outlined in Figure \ref{overview of data models}, the web interface guides users to author policies as illustrated in Figure \ref{web interface}: Steps 1 and 2 showcase the configuration interface of the \term{custom action} and \term{base procedure}. At steps 3 and 4, users will then be guided to specify the \term{decorators} and \term{executions} respectively.  

At each step of the authoring process, the left \leijie{side panel} (Figure \ref{web interface}.A) displays a selection of available options along with detailed explanations. Additionally, to enhance system transparency, users with a basic understanding of programming can also view the source code of the selected option (Figure \ref{web interface}.E). In the right panel, users are asked to configure the \term{settings} of the selected option (\ref{web interface}.B). According to each \term{setting}'s data type, we display various input boxes with validations in place. For \term{settings} corresponding to a specific entity, a drop-down list replaces the standard input box. For example, for the \code{Vote Channel} \term{setting}, rather than requiring users to input the channel ID, we pre-populate a drop-down list with Slack channels fetched from the user's community, allowing for direct selection (\ref{web interface}.C).

As we use \term{variables} to connect policy components in our declarative language, it is important to make sure that end-users can understand and use them easily. We maintain a global list of \term{variables} to tracks them throughout an authoring process. When users start to configure a new \term{setting}, we automatically identify and present relevant \term{variables} from the global list--matching them based on their respective entity and data types--within a drop-down list for selection. For instance, within the \code{Vote Channel} \term{setting} drop-down list (\ref{web interface}.C), users can find an entry named \code{the channel where the message was renamed}, a \term{variable} derived from the \term{base action}. For \term{settings} of the entity \code{Text}, we noticed that users may want to reference both \term{settings} and \term{variables} of diverse data types and entities--for instance, introducing the configuration of a voting procedure in a message sent to eligible voters. Therefore, we've integrated an ``Insert Variables'' button, granting users access to the comprehensive list of \term{variables} and \term{settings} (\ref{web interface}.D).

\begin{figure*}[!h]
    \centering
    \includegraphics[width=\textwidth]{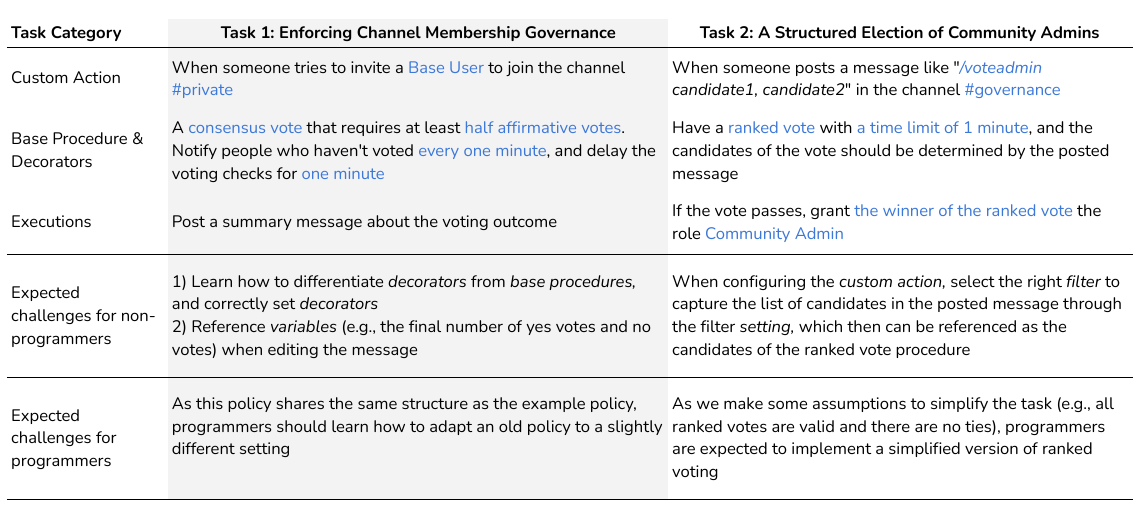}
    \caption{\textbf{Outlines of Authoring Tasks for Non-Programmers}. \textmd{During the user study, we presented a community governance scenario that motivates the creation of each policy. Participants were required to parse out the \term{custom action}, \term{custom procedure}, and \term{executions} of each policy from the scenario and articulate them in our system \System{}.}}
    \hfill
    \label{study tasks}
\end{figure*}

\subsection{System Implementation}
We developed our system within PolicyKit's Django framework. Our system, hosted on a web server, consists of a frontend web interface (JavaScript, HTML, CSS), a Python backend, and a JSON library for policy components. The backend loads all policy components from the JSON library and fetches community information (e.g., slack channels, community users) via platform API. The web interface then guides users to author a policy in the declarative language. Upon the completion of policy authoring, the web interface gathers the newly authored policy in JSON format, which will then be compiled into executable code compatible with PolicyKit.

\section{Evaluation}
In our evaluation, we examined whether \System{} could be learned and used by community members to author a variety of policies in a short amount of time. 
Specifically, we aim to address the following three research questions through two user studies. The first study with non-programmer community members primarily assesses \System{}'s usability and expressivity. The second study with programmers, on the other hand, evaluates its efficiency compared to a code-based baseline for authoring governance policies, PolicyKit~\cite{zhang2020policykit}. We developed our final study protocol iteratively through three pilot interviews with lab members to ensure its effectiveness.

\begin{itemize}
    \item \textbf{Learnability and Usability}. Can non-programmers use our system to author governance policies? 

    \item \textbf{Efficiency}. Can programmers use our system to author policies faster than writing code?

    \item \textbf{Expressivity}. Can non-programmers use our system to author most policies they want to enact?

\end{itemize}

\subsection{For Non-Programmers: Authoring Policies with \System{}}

\subsubsection{\textbf{Recruitment and Participants}} We recruited 10 participants, experienced in community governance on online platforms and not programmers by publishing a call for participation on mailing lists related to community governance. The studies were conducted via video calls, averaging a duration of 79 minutes each, with participants receiving a compensation of \$20.

\subsubsection{\textbf{Study procedure}} We started our user study with an onboarding session. We first introduced to participants the fundamental concepts of \System{} (e.g., \term{policy}, \term{action}, \term{procedure}, \term{execution}) through policy examples. We then gave them a tutorial on how to use \System{} to author a simple policy, \textit{a consensus vote for channel renames}, step by step. This process took 18 minutes on average. 
Participants were then tasked with authoring policies of incremental difficulty based on two distinct community governance scenarios, as illustrated in Figure \ref{study tasks}. The first task, titled \textit{Enforcing Channel Membership Governance}, pertains to a community's need for a consensus vote whenever someone invites a new user to join their private channel. It has a similar structure to the example policy in our tutorial to help participants become familiar with \System{}. The second task \textit{A Structured Election of Community Admins} is more challenging: a community wants to trigger a ranked vote election by posting a message like \textit{\%voteadmin candidate1, candidate2, candidate3}. As such command-triggered policies are often used in community governance tools~\cite{slackcommand}, it is important to ensure non-programmers can author this kind of policies in our system. 

Participants were asked to speak aloud their thoughts and confusion as they worked. Researchers were silent except to clarify the details of task scenarios and the web interface. If they spent over 5 minutes on a subtask but were not close to succeeding, the researchers would offer hints or explain the answer and mark the respective task as failed. We measured the time participants spent on each task and asked them to fill in a post-survey that assesses the task workload and system usability through Task Load Index~\cite{hart1988development} and System Usability Scale~\cite{brooke1996sus} respectively. 

To gain preliminary insights into our system's expressivity, we asked each participant what governance policies their communities have enforced (either manually or algorithmically) and what kinds of policies they envision authoring using \System{} at the end of the study. We will have more detailed discussion about the study process and results of expressivity evaluation in Section \ref{expressivity}.

\subsubsection{\textbf{Results}}
9 out of 10 participants successfully completed the first task, and 7 out of 9 successfully completed the second task. One participant had to leave early so did not take part in the second task. On average, participants who succeeded took 7.5 minutes and 8.6 minutes for two tasks respectively. Most participants could quickly learn how to articulate a policy in our system. While the participant who failed in the first task had trouble navigating the \System{}, the two participants who failed in the second task struggled with understanding command-triggered policies. Even participants who succeeded also spent a significant amount of time learning how to connect the list of community users extracted from the triggering message to the candidates of the governing procedure. 

We asked participants to rate the system's usability via a post-survey. On a 5-point Likert scale (1--strongly disagree, 5--strongly agree), participants rated 3.90 on average on the statement ``I think I would like to use this system frequently'' and 2.30 on the statement ``I would need a technical person to be able to use this system''. However, participants also acknowledged they should spend some effort learning how to use this system, as evidenced by the rating of 2.90 for ``I would still need to learn a lot before using this system,'' and 2.80 for ``I thought the system was easy to use''.

\begin{figure}[!ht]
    \centering
    \includegraphics[width=\columnwidth]{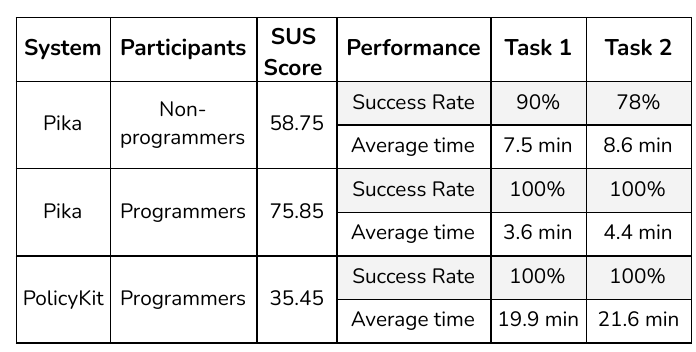}
    \caption{\textbf{Results of Two User Studies}. \textmd{We calculated the average times after excluding data from participants who failed to complete the tasks. We observed that most non-programmer participants can use \System{} to articulate policies around 2.5 times faster than programmers across two different governance scenarios. We also reported the usability score measured by the System Usability Scale~\cite{brooke1996sus}. We found that \System{}  vastly outperformed PolicyKit from the perspective of programmers, and was also rated as usable by non-programmers.}}
    \hfill
    \label{study results}
\end{figure}

\subsubsection{\textbf{Common Mistakes and Observations}}

Participants primarily struggled with articulating the \term{custom action} in the second task. It proved challenging as the election candidates were derived from the triggering message---this concept of command parameters confounded several participants. Participants were expected to choose a specific \term{filter} \code{only texts that begin with a command and are followed by a list of community users} for the \code{message} field of the \term{base action}. This choice allows the list of candidates to be captured as \term{variables} and available for references in the \term{base procedure}. However, participants often selected the \term{filter} \code{only texts that start with the specified word} but only changed it upon realizing they could not set the election candidates later. While this challenge partly arose from non-programmers' unfamiliarity with command parameters, a real-time feedback system that displays what kind of actions are now governed could be beneficial.

Many participants liked the global data list that enables them to reference \term{variables} across components easily and create a more generic policy (e.g., a policy that governs any renaming channel actions through a voting procedure in the channel subject to renaming). However, we also noticed participants often tended not to use \term{variables} whenever possible. For instance, in the second task, as the channel where the triggering message is posted is always the channel \code{\#governance}, almost all participants selected the channel \code{\#governance} directly.

These \term{variables} and \term{settings} also enable participants to deliver more nuanced messages when a procedure happens, passes, or fails. For instance, they wanted to provide a detailed summary of voting outcomes or to warmly welcome a new channel member. Interestingly, 4 out of 10 participants dedicated half of their authoring time solely to fine-tuning these messages. This behavior underscores a shift of focus from the technical aspects of policy authoring to clear communication with community members. As existing research has highlighted the significance of notification messages in community governance~\cite{geiger2012defense}, our study further indicates that non-programmers, when equipped with the appropriate system, can introduce fresh perspectives into the policy authoring process.


\subsection{For Programmers: Authoring Policies with \System{}}

\subsubsection{Recruitment and Participants} We recruited 7 programmers who are interested in community governance. As we aimed to evaluate the efficiency of our \System{}, prior experience in community governance was not a prerequisite for participation. \leijie{Similarly, familiarity with PolicyKit was not a prerequisite given there are only a few communities actively using PolicyKit.} We recruited participants by publishing a call for participation on mailing lists related to community governance and Slack channels of the university CS department. \leijie{All 7 participants are familiar with Python programming and 4 of them have a basic knowledge of PolicyKit}. The user studies were conducted via video calls, with each session averaging a duration of 95 minutes. Participants were compensated with \$40.


\subsubsection{\textbf{Study Procedures}} The second user study is designed to evaluate the efficiency of \System{}. \leijie{We chose PolicyKit~\cite{zhang2020policykit} as our baseline for the following reasons. Our literature review revealed that, aside from PolicyKit, there are limited options for programmers to author comprehensive sets of policies. While most online platforms offer tutorials for bot creation, these often demand extensive knowledge of the platform's software infrastructure and server deployment skills. End-user systems like AutoMod on Reddit support only a limited range of policies, typically for content moderation. While authoring policies in PolicyKit requires familiarity with its documentation, it is still a better baseline than directly implementing bots via platform APIs. In the following, we describe each condition in more details.}


\textbf{System Condition}. The system condition for \System{} mirrors the non-programmers' study: programmers started with a \System{} tutorial and then authored policies based on two distinct community governance scenarios. 

\textbf{Baseline Condition}. We first introduced to participants the basic syntax of PolicyKit and walked them through the implementation of the policy \textit{a consensus vote for channel renames}. We provided comprehensive API documentation detailing relevant functions and classes. Then participants were asked to author a policy given one randomly selected governance scenario from Figure \ref{study tasks}. We limited this task to only one policy to prevent the study from becoming overly time-consuming. We evaluated participants' performance via a close examination of their code. 

Recognizing the difficulty of authoring policies in PolicyKit, we added several facilitating conditions to the baseline condition to make two conditions comparable. Instead of randomizing the order of the two test conditions, we consistently began with the system condition before the baseline condition. This sequence allowed participants to familiarize themselves with crucial concepts before tackling more demanding programming tasks. We also made several simplification in the programming task: for instance, we assumed that all ranked votes are valid, thereby eliminating the need to check the validity of votes. Moreover, participants were allowed to copy the code of the example policy and were exempted from debugging. These arrangements together make our baseline condition a more robust comparison to authoring policies in \System{}.

\subsubsection{\textbf{Results}}
For the system condition, all 6 participants successfully authored two policies using \System{} with an average time of 3.6 minutes and 4.4 minutes. In comparison, for the baseline condition, while all participants were able to author the given policy using PolicyKit, they spent considerably longer time, 19.9 minutes on average for the first task and 21.6 minutes for the second (paired t-test, $p<0.01$). During the study, participants spent much time understanding the workflow of PolicyKit engines and looking for relevant functions in the API documentations. The efficiency of \System{} becomes even more apparent when considering that participants were not required to debug their implemented policies, a task that might have substantially increased their time usage. 
 
We asked participants to rate the usability of both \System{} and PolicyKit via a post-survey. For the Task Load Index survey using a 7-point Likert scale (1--very low, 7--high), participants reported experiencing a significant higher mental workload while using PolicyKit, with average scores being 2.66 points higher for mental demand (paired t-test, $p < 0.05$) and 2.17 points higher for effort level (paired t-test, $p < 0.01$) than \System{}. Regarding the System Usability Score (SUS), \System{} vastly outperformed PolicyKit. The average SUS scores were 75.85 for \System{}, indicating good usability, and 35.45 for PolicyKit that represents less acceptable user experience. A paired t-test revealed a statistically significant difference between the SUS scores of these two conditions ($p<0.01$). These findings demonstrate that, compared to PolicyKit, \System{} significantly improves the ease of authoring governance policies for programmers.

\begin{figure}[!ht]
    \centering
    \includegraphics[width=\columnwidth]{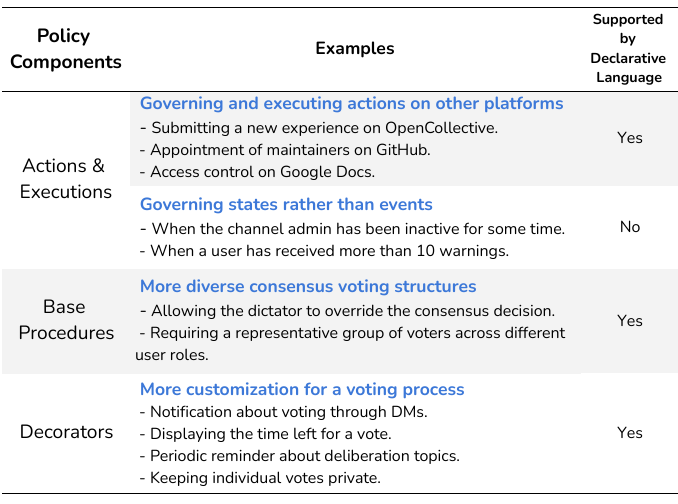}
    \caption{\textbf{New Policy Components Participants Want}. \textmd{During the study, we asked non-programmer participants about policies they enforced or desired in their communities. Here we list representative examples of new policy components they proposed. As a significant number of policy components participants mentioned have already been supported in our library, we do not include them in the table. Instead, refer to Section \ref{libraryofcomponents} for more details.}}
    \hfill
    \label{study tasks}
\end{figure}

\subsection{Expressivity of \System{}}
\label{expressivity}
To gain an understanding of \System{}'s expressivity, we inquired non-programmer community members about policies they enforced or desired in their communities. \leijie{If our library of policy components supported their needs, participants were asked to author these new policies. In cases where their proposed policies required new policy components or needed integration with other online platforms, participants were instead asked to describe these policies using our declarative language. This process helps us understand (1) how our declarative language can support real-world policy authoring, and (2) how non-programmers can use our declarative language to author policies they desire beyond predefined scenarios.} Participants proposed more than 25 policies in total. 

\subsubsection{Expressivity of Declarative Language}
We want to first differentiate between the expressivity of our declarative language and of our library of policy components. As we have only implemented an initial library of policy components, some policies that would fit well into our declarative language still require authoring new components or supporting online platforms other than Slack. As we will discuss in \ref{author_components}, a crowdsourcing approach is crucial to enriching this library. We will prioritize the expressivity of our declarative language in this work.

We found that a significant number of proposed policies are supported by our library of components. Many policies govern a variety of community actions through a voting process. These actions span from platform-specific ones, such as regulating community or channel memberships and message posting, to more high-level actions, such as adding new rules to community documents, appointing admins or moderators, or determining action item priorities. For instance, one policy states that \textit{if a message posted in the \code{\#announcement} channel gathers more than five thumb-down emoji votes, it should be deleted with a warning sent to the message sender}. In addition, participants also wanted to automate some governance tasks without a voting process. Examples include \textit{mandating the invitation of a moderator when a new channel is created} or \textit{having the command \code{!mods} automatically mention all moderators}.

Other policies can be described by our declarative language but require the authoring of new policy components. 
Most prominently, participants desired governing actions on diverse platforms, such as expense submissions on OpenCollective, appointments of maintainers or contributors on GitHub, or access control on Google Docs. As \System{} relies on PolicyKit for action listening and execution, governing actions on other platforms requires the development of the corresponding platform integrations, which is a straightforward process in PolicyKit.
For governing procedures, some participants also envisioned other voting procedures, such as allowing the dictator to override the consensus decision or requiring representatives from different user groups to vote.
In addition, participants also mentioned more \term{decorators}, including voting notification through DMs, displaying the time left for a vote, periodic reminders about deliberation topics, and keeping individual votes private.

However, there exists a small set of policies that \System{} cannot support for now. This limitation arises from \System{}'s focus on event-driven rather than state-driven governance.
Huang~\cite{huang2015supporting} differentiates between triggers based on events (as instantaneous signals) and states (as Boolean conditions that can be evaluated to be true or false at any time) in trigger-action programming. Participants mentioned some policies that govern states: \textit{if the channel admin has been inactive for one month, hold an election for new admins}, or \textit{if a user has more than 10 warnings, kick the user out of the community}. As PolicyKit listens to platform actions through webhooks, we only support policies triggered by event-driven actions. In the future, we plan to set up periodic backend polls to check state triggers.

\subsubsection{Learnability and Usability of Declarative Language beyond Predefined Tasks}
\leijie{We found that most non-programmer participants were able to effectively use Pika’s language to map out the policy components for policies they proposed. However, we identified two main challenges. First, while some state triggers can be reframed as actions about state changes (e.g., when a user has received the 10th warning), participants found state expressions more familiar and natural. Second, there was confusion around whether certain functionalities should be categorized as \term{decorators} or \term{base procedures}. For instance, in a policy that allows a dictator to override a consensus decision, it was unclear whether this required a new procedure (a mixture of dictatorship and consensus voting) or a new decorator (e.g., giving certain individuals higher voting weights). }

\subsection{Feedback from Participants}
The overall reactions to \System{} ranged from positive to enthusiastic.
One non-programmer said: ``\textit{I'm very impressed by the solid basis you have developed, and by how helpful, inviting, and urgently needed, I think it will prove helpful for many online communities and their administrators.}''
Along similar lines, another programmer commented, \textit{``I would say you're ready to release it to users to see how that goes.''}

Many non-programmer participants felt that \System{} would grant them greater agency in community governance. 
One user appreciated the exposure to an extensive array of governance possibilities: ``\textit{Programmers and technical people know what their set of options are so they can start imagining it. This system is really interesting because it starts exposing what options are available that can be played with, what are the things that are captured or not captured in the platform.}'' 
Another participant believed this system can motivate community administrators to document policies: ``\textit{[This new policy] is a nice idea, but I'm not going to document it because I don't have a way to enforce it...Why bother creating that out here? [But] this [system] forces our hand as administrators to start saying like, okay well, what are the rules that I would like to have and then see if I can implement them? Because now there's a way to potentially automate that.}'' 

Even if their community has already had technical developers, participants were enthusiastic about involving non-programmers in the governance process. 
One participant believed \System{} can be used as a proof of concept to facilitate their communication with developers: \textit{``The developers are not usually the ones doing governance. [To ask programmers to develop a policy], you need to have the UI available and easily accessible to people to see what it does. Otherwise, you're only going to get developers hearing about this.''} 
More broadly, participants appreciated the value of incorporating non-programmers' perspective in community governance: ``\textit{I think it's helpful to have a governance panel: there are a coder, the governance person, a layman member of the community who are someone who's like a subject to the rules.}''







\section{Discussion}
\System{} offers a declarative language to articulate governance in online communities. From our evaluations, we found that \System{} can both express a wide range of desired governance policies and is usable to non-programmers and programmers alike. Below, we will describe how \System{} unlocks a range of higher-level extensions that could be built on top of it and enables new research directions.

\subsection{Generalization to Additional Online Platforms}
\leijie{While we built our system prototype around Slack, our design of the declarative language is not tied to any specific platform and can be easily generalized to other platforms. For example, our conceptualization of base procedures requires only a space for group discussion (e.g., posts, channels, or threads) and a method for response (e.g., emoji reactions, threaded replies, or up-vote/down-vote mechanisms). Extending \System{}'s functionality across different platforms is a straightforward process. First, as \System{} relies on PolicyKit platform integrations for action listening and execution, it requires the development of the corresponding platform integrations. Once any developer has written a PolicyKit integration, every community on that platform has the ability to install \System{}. Due to frequent changes in API standards across platforms, we acknowledge that this still requires dedicated engineering efforts; however, \System{}'s design would not need to chance to accommodate the same policies on additional platforms.}

\leijie{Second, our initial library developed for Slack offers a valuable starting point, as many governance procedures are platform-agnostic and can be easily adapted. For instance, as long as the underlying platform integrations use consistently named functions for sending notifications, initiating votes, and enumerating vote outcomes, our implemented voting procedures can be adapted to other platforms via minor changes. Indeed, PolicyKit exposes a unified API to policy authors via the Metagov Gateway library \footnote{\url{https://github.com/metagov/gateway}}, which implements the same named functions across different platform integrations. Drawing parallels between platform features---like Slack's workspaces and Discord's servers, or Slack's messages and Reddit's posts---can further facilitate this adaptation process.}

\subsection{Supporting Deliberative and Participatory Community Governance}

\System{} supports a transition from a programming-centric to a governance design-centric perspective, which paves the way towards more \textit{deliberative} and \textit{participatory} community governance. First, non-programmer community administrators can channel their energy into deliberating their governance policies rather than struggling with coding them. For instance, instead of searching for bots that enable more sophisticated voting processes, communities can focus on discussing which voting process can best reflect community consensus. Similarly, they can spend more time designing  detailed and transparent procedural messages to better communicate a policy,  a critical aspect of community governance often overlooked from a programming perspective~\cite{jhaver2019does}. Developers can also use \System{} to rapidly prototype governance policies.

Second, for non-programmers not involved in designing governance policies, \System{} also empowers them to engage in governance discourse. Given that non-programmers constitute the majority of online users today~\cite{nielesen2016computer}, it is critical to invite them to audit governance policies embedded in code. For instance, Geiger et al. documented that Wikipedians expressed discontent with a bot developed based on inaccurate assumptions regarding social norms and demanded the implementation of an opt-out mechanism~\cite{lovink2012critical}. By translating governance policies in code to a more accessible declarative language, \System{} facilitates a clearer understanding of policies among non-programmers, thereby encouraging them to voice their concerns more effectively. 

In the long term, our ambition extends to creating an ``app store'' for governance~\cite{schneider2022admins} which offers communities a variety of governance models beyond the default ones. As the center of this app store, \System{} enables communities to consistently evolve their governance. Hence, it becomes feasible to analyze which policies---and their specific conﬁgurations---are frequently used. This then opens avenues for comparing governance policies across different communities, leading to the development of more community-tailored policy components. 
\leijie{In addition, such an app store could also motivate community users who have programming expertise to author policy components, as it serves as a dynamic space where programmers of policy components can exchange ideas, receive feedback, and see the real-world impact of their work. }

\subsection{Towards Empowering Programmers to Author Policy Components}
\label{author_components}
In this work, we prioritized designing an expressive declarative language over building a comprehensive list of policy components. As a result, we only implemented a library of components as a starting point. Our user study has indicated that this library can accommodate a significant number of policies participants wanted. However, to truly unlock \System{}'s full potential, crowdsourcing efforts are needed to enrich this shared library.
While programming skills remain required for authoring policy components, developers no longer need to implement a policy from scratch but instead can concentrate on building individual components. 
More importantly, \System{} facilitates the reuse of policy components across communities. When a new component is added to the library, it becomes readily accessible to other communities. This significantly reduces collective efforts in implementing repetitive governance policies.

However, the current version of \System{} only supports adding new components through direct edits to the JSON-formatted library. 
\leijie{This process presents several challenges for programmers. First, programmers might have difficulties in following the specific schema of policy components we enforce to ensure end-user accessibility. Second, coding policy components within PolicyKit is a complex task. It requires not only familiarity with PolicyKit's documentation but also navigating its limited debugging capabilities. In our study, our programmer participants typically spent about 30 minutes learning PolicyKit grammar and implementing a simplified ranked voting procedure. Enhancing the user-friendliness of these policy components can be more time-consuming, which may involve adding default values and validation for \term{settings} and sending procedural messages (e.g., when they cast invalid votes).}

To overcome these challenges, we envision introducing a form-based web interface that guides programmers through authoring new components. This interface would ask programmers to specify a component's descriptive attributes, determine component \term{settings} that require user input, expose \term{variables} to connect with other components, and finally provide the corresponding code snippet. 
This interface could also help programmers author code snippets that adhere to the required format.
Additional tools for debugging, simulation, and testing are also essential to facilitate the component authoring process. Given a component needs to be integrated into a policy to fully operate, this interface should simulate the enforcement of the whole policy, allowing programmers to observe the effectiveness of their authored components. 

\subsection{End-User Programming in Community Governance}
During the design process, we made deliberate design choices to tailor \System{} to the needs of non-programmers. At times, this meant sacrificing the expressivity of some aspects of the system. 
For instance, while we envisioned a generic \term{filter} that captures a command followed by parameters of any type or entity, we realized that this \term{filter} would make non-programmers grapple with determining whether a generic parameter can be used for a specific component \term{setting}. Therefore, we opted for a more restricted \term{filter} for each use case---for instance, one that starts with a command and is followed by a list of community users. 
Similarly, to simplify the authoring process for non-programmers, we categorize \term{executions} that happen when the procedure starts (e.g., notifying moderators about the start of a voting procedure) as \term{decorators} but not as \term{executions} in our declarative language.

While we primarily drew design inspiration from trigger-action programming~\cite{dey2006icap, ur2014practical}, non-programmers can also benefit from other techniques widely used in end-user programming research~\cite{myers2006invited, barricelli2019end}.
For instance, a visual programming interface with drag-and-drop components might offer them a more intuitive grasp of policy flows than \System{}'s form-based authoring interface~\cite{akiki2017visual, dorner2011supporting}. 
Real-time feedback about the expected policy behaviors based on users' selection can also aid non-programmers in debugging their policies. 
Moreover, similar to how people build upon existing recipes in trigger-action programming tools~\cite{ur2014practical}, \System{} could offer a series of representative policies, enabling users to modify these examples as an initial step towards mastering their policy authoring.

Although our declarative language is designed to accommodate a broad spectrum of policies, certain communities might need a specific kind of policies and therefore find an expressive declarative language unnecessarily complex. For instance, financial management platforms (e.g., OpenCollective) might only need policies regarding the categorization and approval process of different expenses. In such scenarios, we envision our declarative language serves as a foundational framework, upon which communities may build custom apps that restrict the expressivity of our declarative language or selectively use components from the library. These custom apps can communicate seamlessly with PolicyKit backend via our declarative language. These apps may also have a simplified authoring interface that integrates with platform clients, enhancing their usability for non-programmer community members.


\section{Limitation and Future Work}
\leijie{
As we focus on empowering non-programmers to author governance policies that are commonly desired, we do not explore explore how \System{} can engage community members in making ``good'' policies. 
This limitation arises from the inherent challenge of defining ``good'' governance, which may vary greatly across communities~\cite{fiesler2018reddit, wang2022ml}.
To mitigate this limitation, we have curated a library of policy components, based on what we identified as ``best practices'' in community governance from prior literature and \revise{our observations from deploying PolicyKit with a few communities for several years}. We also focus on empowering knowledgeable community members to \textit{easily} and \textit{expressly} author policies they want. Furthermore, achieving good governance is often an experiential and learning process~\cite{de2007governance}. \System{} can support communities to iteratively develop and refine governance policies. Looking forward, future work could uncover guidelines on designing ``good'' policies for a given community or provide tools to comprehensively evaluate policies. Other future longitudinal field studies with real-world online communities are also necessary to fully understand Pika's impact on community policymaking processes.}


\leijie{We also envision having a fine-grained classification inside each category of policy components. For instance, voting procedures share a similar set of \term{settings} like eligible voters and voting channels, and can be considered as a special group of \term{base procedures}. Correspondingly, there are a group of \term{decorators} that are only compatible with these voting procedures, such as notifying voters. In our current implementation, we standardize the naming of shared \term{settings} across voting procedures, and ensure that the corresponding \term{decorators} reference these standardized names. As we expand our library with additional \term{base procedures} and \term{decorators}, we aim to clearly define the interfaces for each category of \term{base procedures}. This will not only ensure that newly added procedures are compatible with existing \term{decorators}, but also help non-programmers determine whether a \term{decorator} is applicable to a \term{base procedure}}.

\leijie{In addition, as our user studies only illustrate the usability and expressivity of \System{} in a controlled lab setting, future longitudinal field studies with real-world online communities should also help further uncover new usability and expressivity issues. Finally, we also plan to support a range of online platforms beyond Slack by developing PolicyKit's platform integrations and creating a library of policy components. }

\section{Conclusion}
In this work, we present \System{}, a novel system that empowers non-programmers to create a broad range of governance policies that are directly executable on their home platforms. Our system consists of three parts. At its core, \System{} uses a declarative language that decomposes governance into modular policy components, thereby scaffolding an expressive policy authoring process for non-programmers. The modular design also makes it possible to have a library of policy components and flexibly add new ones. We implemented a library of policy components to support the authoring of a number of common policies used by online communities. Finally, we built a user-friendly, form-based web interface to guide non-programmers to author policies based on this declarative language and policy component library. Our user studies with 10 non-programmers and 7 programmers show that \System{} can empower non-programmers to author governance policies around 2.5 times faster than programmers who author in code. We also provide insights about \System{}'s expressivity in supporting an extensive array of policies that online communities want.


\begin{acks}
\end{acks}


\bibliographystyle{ACM-Reference-Format}
\bibliography{policykit}


\end{document}